# Single quantum dot nanowire LEDs

*Ethan D. Minot\*, Freek Kelkensberg, Maarten van Kouwen, Jorden A. van Dam, Leo P. Kouwenhoven*

*and Valery Zwiller*

Kavli Institute of Nanoscience, Delft, The Netherlands

*Magnus T. Borgström, Olaf Wunnicke, Marcel A. Verheijen and Erik P. A. M. Bakkers*

Philips Research Labs, Eindhoven, The Netherlands

\* Corresponding author. E-mail: ethan@qt.tn.tudelft.nl

We report reproducible fabrication of InP-InAsP nanowire light emitting diodes in which electron-hole recombination is restricted to a quantum-dot-sized InAsP section. The nanowire geometry naturally self-aligns the quantum dot with the *n*-InP and *p*-InP ends of the wire, making these devices promising candidates for electrically-driven quantum optics experiments. We have investigated the operation of these nano-LEDs with a consistent series of experiments at room temperature and at 10 K, demonstrating the potential of this system for single photon applications.

Nanowire light emitting diodes (NW LEDs) offer exciting new possibilities for opto-electronic devices. Growth of direct-bandgap NWs on Si[1, 2] will allow optically active elements to be integrated with already highly mature Si technology. For solid-state lighting applications, broad-area LEDs made from NW arrays have higher light-extraction efficiency than traditional planar LEDs[3], and in the field of quantum optics, NWs offer the possibility to control electron transport at the single-electron level[4] and light emission at the single-photon level[5].

Since the first demonstration of GaAs NW LEDs in 1992 [6], different geometries and materials have been used to produce NW LEDs operating over a wide range of wavelengths[3, 7-10]. Single-NW LEDs with doping modulation in the axial direction, which is the most interesting geometry for many applications, have been fabricated using GaN-GaInN multi-junctions[3] and a proof-of-principle device has been shown using InP[7]. In this letter we describe the fabrication and characterization of reproducible axial InP NW LED devices, and show that an active InAsP quantum dot region can be incorporated into these devices. The axial geometry allows for controllable injection of electrons and holes into the precisely defined active region, with the additional advantage of high light-extraction efficiency since the optically active region is not embedded in a high refractive index material. Unlike GaInN, InAsP emission can be tuned to infra-red telecommunications wavelengths where there is strong interest in electrically driven single-photon sources[11].

Nanowire *p-n* junctions were reproducibly grown in the vapor-liquid-solid (VLS) growth mode[12] by use of low-pressure metal-organic vapour-phase epitaxy (MOVPE). 20 nm colloidal Au particles were dispersed on (111)B InP substrates, after which the samples were transferred to a MOVPE system (Aixtron 200), and placed on a RF-heated gas foil rotated graphite disc on a graphite susceptor. The samples were heated to a growth temperature of 420 °C under phosphine ($PH_3$) containing ambient at molar fraction $\chi_{PH3} = 8.3\times10^{-3}$, using hydrogen as carrier gas (6 l/min $H_2$ at 50 mbar). After a 30 s temperature stabilization step, the NW growth was initiated by introducing trimethyl-indium (TMI) into the reactor cell at a molar fraction of $\chi_{TMI} = 2.2\times10^{-5}$. During the first 20 minutes, hydrogen sulfide ($\chi_{H2S} = 1.7\times10^{-6}$) was used for *n*-type doping, after which the *p*-type NW part was grown by switching to diethyl-zinc ($\chi_{DEZn} = 3.6\times10^{-4}$) as a dopant source for another 20 minutes of NW growth. A low growth temperature of 420 °C was used to minimize thermally activated radial overgrowth[13] that will degrade the electrical properties of axial modulation doping[14]. To incorporate nominally intrinsic $InAs_xP_{1-x}$ heterostructure segments in the InP *p-n* junctions, both dopant gases were shut off and arsine ($AsH_3$) was added to the gas flow between *n*- and *p*-type InP growth. The fraction of As in the $InAs_xP_{1-x}$ segment was controlled by adjusting the $\chi_{AsH3} / \chi_{PH3}$ ratio. After growth, PH$_3$ was kept present during cooling.

A typical NW LED device without an InAsP heterostructure is shown in Figure 1a. A total of 20 such devices have been successfully fabricated with a process yield of about 50 %. As-grown NWs were transferred to oxide coated Si substrates and contacted using electron beam lithography and metal evaporation. Immediately before metal evaporation, an HF etch was used to remove native oxide from the NW surface (5 seconds in 6:1 buffered HF). Ti/Al (100/10 nm) was used to contact n-InP and Ti/Zn/Au (1.5/30/90 nm) was used to contact p-InP. The contact properties, similar to those previously reported[7], were determined on individual n-InP and p-InP NWs (see supporting information). Contact to n-InP is Ohmic and typical two-terminal resistance is about 20 kΩ, suggesting a doping concentration > $3 \times 10^{18}$ cm$^{-3}$ (electron mobility in bulk InP is < 1000 cm$^2$V$^{-1}$s$^{-1}$ [15]). Contact to p-InP is non-ohmic, with room-temperature low-bias resistance of about 10 MΩ when the p-electrodes are 1 μm wide. Gating of individual p-type wires shows pinch-off at about 10 V, which suggests a doping concentration of ~ $10^{18}$ cm$^{-3}$ [4]. We note that doping estimates are complicated by surface depletion which reduces the effective diameter of the p-type conductance channel[16], and by surface charges on the NW which can screen the backgate.

The p-InP was observed to be more susceptible to tapering than the n-InP, which can be seen as a diameter expansion on p-side of the junction (Fig. 1a and Fig. 3). Different susceptibility to radial overgrowth is likely due to enhanced materials diffusion on the n-InP surface. The n-type dopant, sulfur, is a well known surface passivant, therefore a thin sulfur protective shell on n-InP is a likely explanation for this effect.

Nanowire p-n junctions were first characterized using three complementary techniques: DC electrical measurements, Kelvin probe force microscopy (KPFM)[17], and photocurrent measurements. Figure 1b shows current as a function of voltage, $I(V_{bias})$, for a NW p-n junction. For small forward bias ($V_{bias} < 2$ V), the current increases following $I \propto \exp(eV_{bias}/nkT)$ where e is the electron charge, k is the Boltzmann constant, T is temperature, and the ideality factor, n, is 4.1. Typical bulk InP diodes have n

≈ 2 due to mid-gap states[18]. The larger $n$ seen in our NW devices may be due to the non-ohmic behavior of the p-type contact. In reverse bias our NW diodes show almost no current (~ 10pA at –2 V) until breakdown occurs at –15V. Kelvin probe force microscopy was used to spatially resolve the electric field within the p-n junction. This technique allows imaging of the electrostatic surface potential in semiconductor devices[19]. Figure 1c shows a NW p-n junction in reverse bias. A drop in surface potential is seen in the middle of the NW, indicating the location and the length, $L$, of the depletion region. From the surface potential data we estimate $L$ to be about 200 nm. Lastly, the p-n junction was characterized by photocurrent measurements (not shown). A 532nm laser was used to excite electron-hole pairs in the NW. Electron-hole pairs created in the depletion region contribute to a photocurrent, driven by the built-in electric field of the junction. At zero $V_{bias}$ we measure 15 pA per Wcm$^{-2}$ which is consistent with a depletion length $L \sim 100$ nm (assuming wire diameter $D = 40$ nm and 40% of photons hitting the area $DL$ are absorbed [20] and converted to current). The size of the depletion region, measured by photocurrent and by KPFM, is consistent with our estimates of n-type and p-type doping concentrations.

Electroluminescence (EL) from the NW LED becomes detectable in forward bias (Fig. 2a and b). At $T = 10$ K we estimate the quantum efficiency (QE) of this device (photons generated/electrons injected) to be about $10^{-4}$ (light detection efficiency of the EL measurement is ~ 1 %). Quantum efficiency varies between devices, with low temperature efficiencies of $10^{-5}$ to $10^{-4}$ and room temperature efficiency about 10 times lower. A significant loss of efficiency is likely due to minority carrier escape to the metal contacts; the minority carrier diffusion length in InP is > 1 μm [15]. There is additional efficiency loss in the p-type segment of the LED; radiative recombination efficiency in Zn-doped p-InP NWs is about 10 times lower than in n-InP NWs [16]. The EL spectrum is significantly blue shifted from the InP bandgap energy (1.42 eV at 10 K). This effect is also seen in heterostructure wires and will be discussed below.

To achieve light emission from an active region with precisely defined dimensions, we have incorporated a nominally intrinsic InAsP section inside the NW p-n junction. The lower-bandgap InAsP material is chosen as a potential well for both electrons and holes, ensuring that recombination will occur in the InAsP, in a similar manner as previously demonstrated for MOVPE grown GaAs-GaP single-photon-emitter NW heterostructures[5].

Figure 3 shows a high resolution TEM image of an InP p-n junction with an embedded InAsP heterostructure segment. In contrast to bulk InAsP, the heterostructure segment has wurtzite crystal structure. The stoichiometry of $InAs_xP_{1-x}$ was determined by electron dispersive x-ray (EDX) TEM measurements with spatial resolution of 3 nm (see supporting information for axial and radial profiles of As content). The axial length of the $InAs_xP_{1-x}$ segment depends on growth time. We used growth times of 50 seconds and 5 seconds to grow 50 nm and 12 nm segments. In the 50 nm segment, shown in Fig. 3, the central region has uniform $x = 0.6$, but As content gradually drops within 20 nm (10 nm) of the n-InP (p-InP) interface. Non-sharp interfaces may be due to As carry-over[21].

The optical properties of the $InAs_xP_{1-x}$ segments in individual NWs were characterized by micro-photoluminescence (micro-PL) spectroscopy. Figure 4 shows emission spectra from three different NWs, each with a different $i$-$InAs_xP_{1-x}$ section sandwiched between $i$-InP ends. By using low laser-excitation powers, emission is detected only from the InAsP. The two spectra with broad peaks centered at 1.17 eV and 1.31 eV are measured from 50 nm long $InAs_xP_{1-x}$ segments with maximum $x = 0.6$ and 0.25 respectively. As expected, higher As content corresponds to lower emission energy. The sharp peak at 1.24 eV is from a 12 nm InAsP segment grown with the maximum $x = 0.5$. The sharp linewidth (1 meV) and blue shift that results from shrinking the heterostructure dimensions is strongly suggestive of quantum confinement.

A similar 12 nm $i$-InAsP section has been incorporated into NW LED devices. Figure 5 shows the EL spectra from such a device. At low current, the spectrum consists of a single peak that includes 1.24 eV emission. As current/voltage increases, this peak grows towards high-energies, but remains below the InP bandgap energy (vertical dashed line). When $V_{bias} > 1.6$ V a second peak emerges in the EL spectrum. This high-energy peak continues to grow as $V_{bias}$ increases, becoming broader and more blue-shifted at large $V_{bias}$.

The low-energy peak is attributed to electron-hole recombination in the InAsP section. The recombination energy is consistent with PL characterization (Fig. 4) and the blue shift of this low energy peak, with increasing current through the junction, is consistent with state filling of the InAsP. These results show that electron-hole recombination can be restricted to a quantum-dot-sized InAsP active region at low currents.

There is a striking difference between the spectral linewidth seen in Fig. 4 (sharp peak at 1.24 eV) and Fig. 5. The PL spectra (Fig. 4) was measured from an undoped NW in which single uncharged excitons can be excited. In contrast, the electrically-driven InAsP section in Fig. 5 will be remotely doped by nearby *p*- or *n*-InP. The broad linewidth seen in Fig. 5 is likely due to overcharging of the InAsP[22] by the surrounding highly doped InP. To further develop the NW LED as an electrically-driven photon source for quantum optics[11, 23, 24], the active region should be shielded by appropriate barriers of intrinsic InP.

The high-energy emission (> 1.42 eV) in Fig. 5 gives insight into transport mechanisms in the LED device. The emergence of this high-energy peak is due to recombination in the InP outside the InAsP heterostructure. The blue shifting and peak broadening at high current indicates that significant state filling occurs in the InP. For a qualitative transport model, we neglect recombination, since minority carrier diffusion length is longer than device length[15]. At 1 µA, the 0.2 eV blue shift corresponds to ~ $10^4$ electrons in the conduction band of a 1 µm section of *p*-InP NW. This electron density and current level is consistent with a minority-carrier electron mobility of ~ 10 $cm^2V^{-1}s^{-1}$. To fully model transport in the NW LED, self-consistent solutions for carrier densities and electrostatic potential must be calculated for both ends of the NW. Our results show that both electrochemical and electrostatic energy play an important role in transport when $V_{bias}$ > 1.6 V.

Using the InP-InAsP system we have demonstrated single-NW LEDs in which carriers are injected into a quantum-dot-sized active region. These results show that InP-InAsP NW LEDs are promising candidates for combining single-electron and single-photon control. Photocurrent measurements show that InP *p-n* junctions can be used for photon detection, opening the way for both photon generation and detection within single NW waveguides[25]. Tuning the properties of the quantum dot active region in the NW LED may enable higher efficiency electrically-driven single-photon sources[11, 23, 24], photon-to-electron turnstiles[26, 27], and novel experiments in solid-state quantum optics, such as spin qubit manipulation followed by conversion to photon polarization state[28].

**Acknowledgment.** We thank Umberto Perinetti for help with PL measurements. AFM image processing was performed with WSxM free software (available from www.nanotec.es). This work was partially supported by FOM, NWO, the Marie Currie program, the European FP6 NODE (015783) project and the ministry of economic affairs of the Netherlands.

**Supporting Information Available:** Current-voltage characteristics of homogeneously *n*-doped and *p*-doped InP NWs (Fig. S1). Current-voltage characteristics of an InP NW *pn* junction at different temperatures (Fig. S2). Electron dispersive x- ray TEM measurements showing axial and radial profiles of elemental composition in 50 nm and 12 nm InAsP heterostructures (Fig. S3).

**Figure 1.** a) AFM height image of a LED device. The *p*-InP/*n*-InP NW is electrically contacted on the left and right side by Ti/Zn/Au and Ti/Al respectively. The surface roughness at the *p*-type electrode is due to Zn grain formation. A step-like change in NW diameter, from 55 nm to 45 nm, corresponds to the location of the *p-n* junction. Scale bar (for x-y dimensions) is 1µm. b) Room temperature *I-V* characteristic of a NW LED device. The inset shows data on a log plot. c) Surface potential map of a device in reverse bias (top) and line trace along NW axis (bottom). The step in surface potential corresponds to the depletion region of the *p-n* junction (highlighted in yellow). The change in NW diameter occurs within this 200 nm depletion region. There is not a one-to-one correspondence between applied voltage and measured surface potential due to capacitance between the AFM tip and the grounded substrate[19].

**Figure 2.** Electroluminescence properties of an InP NW LED without InAsP section. a) Optical microscope image collected by CCD camera of a device at zero bias (left image) and under forward bias (right image). Scale bar is 2 µm. b) Room temperature and 10 K spectrum from the device at $I = 1$ µA. All spectra were collected on a liquid Nitrogen cooled Si-CCD array. c) Integrated light emission as a function of current at $T = 10$ K.

**Figure 3.** Transmission electron microscope (TEM) image of an *n*-InP/*i*-InAs$_{0.6}$P$_{0.4}$/*p*-InP NW. Rotational stacking defects are seen in the InP segments, with a higher density of stacking defects in *n*-InP. The *p*-InP has zincblende crystal structure, whereas the InAsP segment is wurtzite. The *n*-type InP shows a mixture of zinc blende and wurtzite structure. Inset shows a high resolution TEM image of the InAsP crystal structure.

**Figure 4.** Photoluminescence spectra from InAsP sections in three different *i*-InP/ *i*-InAs$_x$P$_{1-x}$/ *i*-InP NWs. The length and As content of the InAsP section has been varied. Red and blue curves are spectra from 50 nm long segments with $x = 0.6$ and 0.25 respectively. The black curve shows emission from a 12 nm long segment with maximum $x = 0.5$. The sharp linewidth of the black curve suggests quantization of energy levels in the InAsP region.

**Figure 5.** a) Electroluminescence spectrum at 10 K from an InP NW LED with a 12 nm InAsP active region. The spectra correspond to 1.4, 1.6, 1.8, 2.0 and 2.2 V bias and 60, 60, 30, 10 and 10 s integration times respectively. Photon counts have been corrected for Si detector sensitivity which drops to zero below 1.1eV. The vertical dashed line indicates the InP bandgap energy at 10 K. Inset shows the *I*-$V_{bias}$ characteristics of the device. The current becomes limited for $V_{bias} > 2$ V. All devices show this behavior which we attribute to hole-depletion near the positively biased contact.

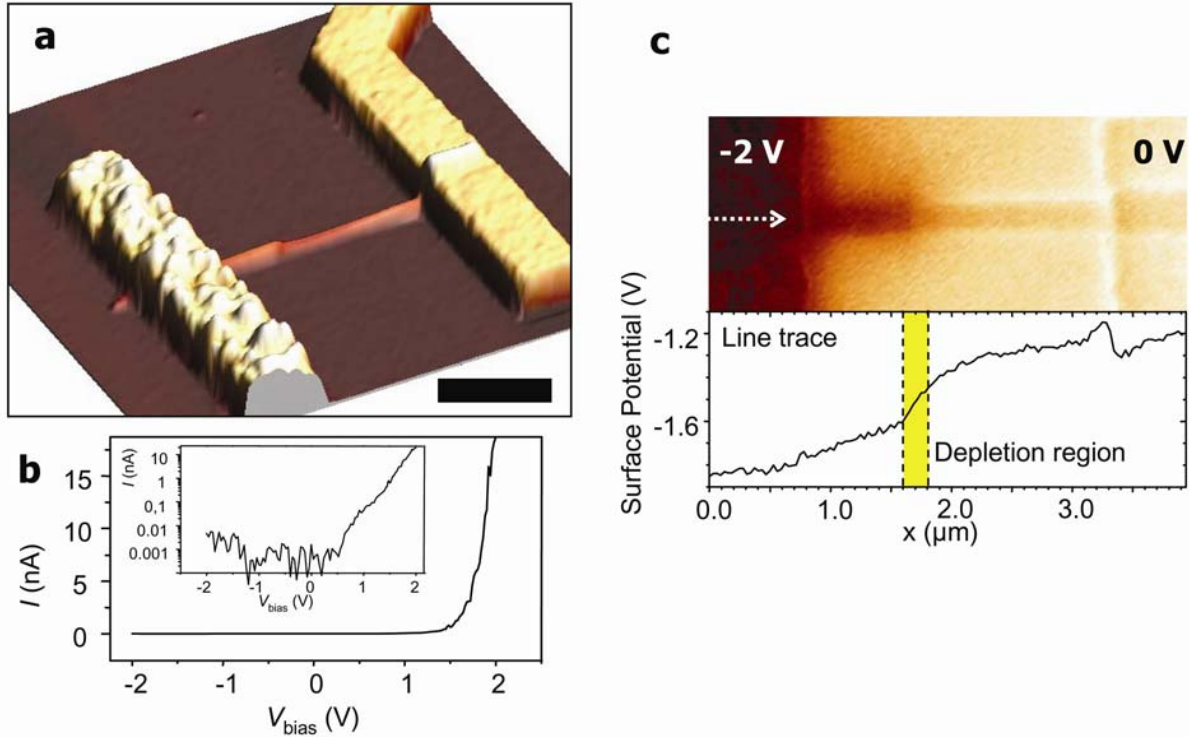

**Figure 1.** a) AFM height image of a LED device. The *p*-InP/*n*-InP NW is electrically contacted on the left and right side by Ti/Zn/Au and Ti/Al respectively. The surface roughness at the *p*-type electrode is due to Zn grain formation. A step-like change in NW diameter, from 55 nm to 45 nm, corresponds to the location of the *p-n* junction. Scale bar (for x-y dimensions) is 1µm. b) Room temperature *I-V* characteristic of a NW LED device. The inset shows data on a log plot. c) Surface potential map of a device in reverse bias (top) and line trace along NW axis (bottom). The step in surface potential corresponds to the depletion region of the *p-n* junction (highlighted in yellow). The change in NW diameter occurs within this 200 nm depletion region. There is not a one-to-one correspondence between applied voltage and measured surface potential due to capacitance between the AFM tip and the grounded substrate[19].

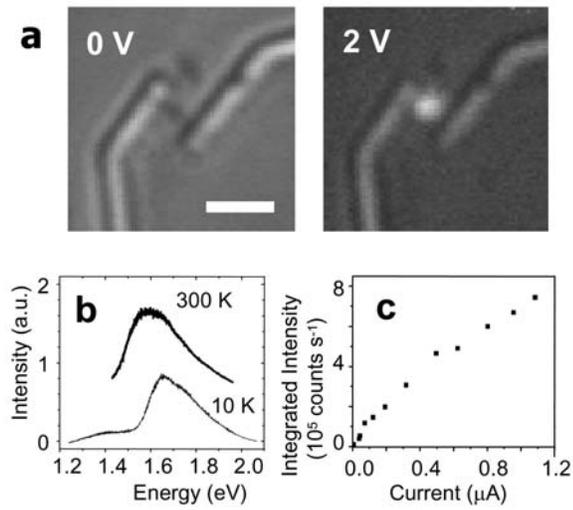

**Figure 2.** Electroluminescence properties of an InP NW LED without InAsP section. a) Optical microscope image collected by CCD camera of a device at zero bias (left image) and under forward bias (right image). Scale bar is 2 µm. b) Room temperature and 10 K spectrum from the device at $I = 1$ µA. All spectra were collected on a liquid Nitrogen cooled Si-CCD array. c) Integrated light emission as a function of current at $T = 10$ K.

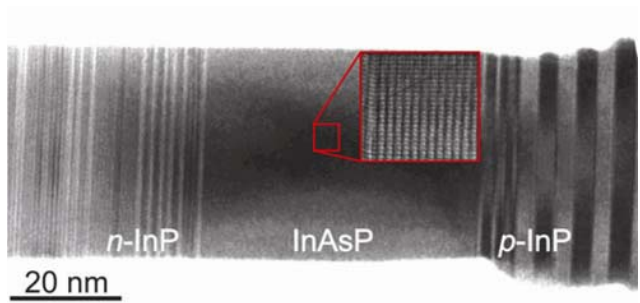

**Figure 3.** Transmission electron microscope (TEM) image of an $n$-InP/$i$-InAs$_{0.6}$P$_{0.4}$/$p$-InP NW. Rotational stacking defects are seen in the InP segments, with a higher density of stacking defects in $n$-InP. The $p$-InP has zincblende crystal structure, whereas the InAsP segment is wurtzite. The $n$-type InP shows a mixture of zinc blende and wurtzite structure. Inset shows a high resolution TEM image of the InAsP crystal structure.

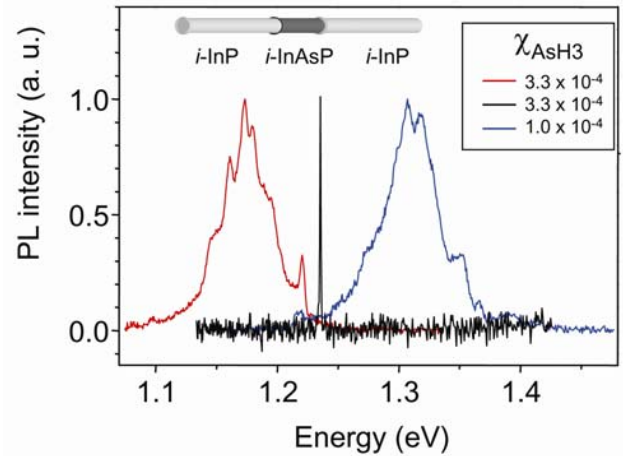

**Figure 4.** Photoluminescence spectra from InAsP sections in three different $i$-InP/ $i$-InAs$_x$P$_{1-x}$/ $i$-InP NWs. The length and As content of the InAsP section has been varied. Red and blue curves are spectra from 50 nm long segments with $x = 0.6$ and 0.25 respectively. The black curve shows emission from a 12 nm long segment with maximum $x = 0.5$. The sharp linewidth of the black curve suggests quantization of energy levels in the InAsP region.

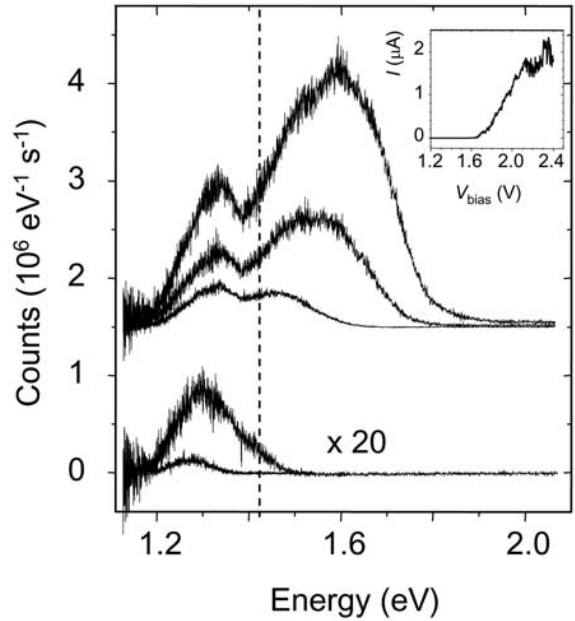

**Figure 5.** a) Electroluminescence spectrum at 10 K from an InP NW LED with a 12 nm InAsP active region. The spectra correspond to 1.4, 1.6, 1.8, 2.0 and 2.2 V bias and 60, 60, 30, 10 and 10 s integration times respectively. Photon counts have been corrected for Si detector sensitivity which drops to zero below 1.1eV. The vertical dashed line indicates the InP bandgap energy at 10 K. Inset shows the $I$-$V_{bias}$ characteristics of the device. The current becomes limited for $V_{bias} > 2$ V. All devices show this behavior which we attribute to hole-depletion near the positively biased contact.